\documentstyle[sprocl]{article}
\def\be{\begin{equation}}
\def\ee{\end{equation}}
\def\bea{\begin{eqnarray}}
\def\eea{\end{eqnarray}}

\def\sb {\mbox{\footnotesize $\Box$}}
\def\sbb {{\mbox{\footnotesize $\Box$}}^\ast}

\def\pq  {P_{\rm Q}}
\def\pl  {P_{\ell}}
\def\pu  {\bar{P}_{\rm U}}
\def\pd  {\bar{P}_{\rm D}}
\def\pn  {\bar{P}_{\rm N}}
\def\pe  {\bar{P}_{\rm E}}

\def\phiu {\bar{\phi}_{\rm u}}
\def\phid {\bar{\phi}_{\rm d}}
\def\Phiu {\bar{\Phi}_{\rm u}}
\def\Phid {\bar{\Phi}_{\rm d}}

\def\Gu {G_{\rm u}}
\def\Gd {G_{\rm d}}
\newcommand{\mbf}[1]{\mbox{\boldmath{$#1$}}}

\begin{document}
\title{Model building in SUSY QCD}
\author{ Naoyuki Haba }
\address{Faculty of Engineering, Mie University, Tsu Mie 514, Japan}
%
%
\maketitle
\abstracts{
In $N=1$ Supersymmetric gauge theories, 
we can calculate non-perturbative effects in the superpotential.
It is interesting to apply this effect to the phenomenology. 
We show two models here. 
One is the low energy supersymmetry (SUSY) breaking scenario 
without messenger sector, 
and the second is the composite model generating 
Yukawa interaction dynamically. 
}

\section{Introduction}

Recently, we can calculate the non-perturbative effects 
in the $N=1$ SUSY gauge theories \cite{1}. 
Many people are trying to apply this effect to 
the phenomenology. 
We will show two models among them. 
One is the low energy SUSY breaking scenario 
without messenger sector, and the second is 
the composite model generating Yukawa couplings 
dynamically. 
\par
We consider $SU(N_c)$ gauge group and 
$N_f$ (ant-)fundamental representations 
$Q, (\overline{Q})$ for the matter. 
The non-perturbative effects, which show vacuum structures, 
largely depend on the number of $N_f$. 
We will show the model building 
in $N_f = N_c -1$ and $N_f = N_c + 1$. 
Each case induces the non-perturbative 
superpotential \footnote{Here $M \equiv Q \overline{Q}$, 
$B \equiv \epsilon Q \cdots Q, \overline{B} \equiv \epsilon
\overline{Q}\cdots \overline{Q}$. 
Since the superfields are boson, 
the gauge invariant operator $B, \overline{B}$ exist 
only in the region $N_c \leq N_f$. }  
\begin{eqnarray}
\label{univ}
   & &  W_{dyn} = {\Lambda^{b_0} \over {\rm det} M}  
    \;\;\;\;\;\;\;\;\;\;\;\;\;\;\;\;\;\;\;\;\;\;\;\;\;\;\;
      ( N_f = N_c -1 ) , \\
\label{1}
   & &  W_{dyn} = {1 \over \Lambda^{b_0}} 
          ( B M \overline{B} -{\rm det} M ) 
    \;\;\;\;\;      ( N_f = N_c+1 )  .  
\end{eqnarray}
%

\section{A model of gauge mediation}

\subsection{Why we need SUSY breaking theory?}

Before showing a model let us consider briefly 
why we need SUSY breaking theory. 
SUSY guarantees the smallness of Higgs mass 
through the chiral symmetry of fermions. 
It is the main motivation of introducing 
SUSY from the viewpoint of phenomenology. 
The standard model (SM) can be extended 
to the supersymmetric standard model (SSM), and 
the minimal extension is so-called 
minimal supersymmetric standard model (MSSM). 
If SUSY is exact, the parameter number 19(1) 
in the SM \footnote{The number in ( ) shows 
the number of $CP$ phase.}$^{,}$
\footnote{Gauge couplings 3; 
quark/lepton masses 9; 
KM matrix 4(1); Higgs mass 1; Higgs self-coupling 1;
$\theta$ 1.} 
does not increase in the MSSM. 
However, since SUSY is broken in our real world, 
the parameter number increases expansionary by 
the SUSY breaking parameters. 
The total parameter number becomes 125(44) 
in the MSSM \footnote{Gauge couplings 3; 
quark/lepton masses 9;
KM matrix 4(1); 
$\theta$ 1; 
$\mu$ 2(1);
$B$ 2(1); 
$m_{H_u}^2$ 1;
$m_{H_d}^2$ 1;
$\xi$ 1;
gaugino mass $M_{\lambda}$ $2(1) \times 3$; 
squark/slepton mass $\tilde{m}^2$ $9(3) \times 5$;
$A$ $18(9) \times 3$;
minus $4$ phases (field redefinition), 
which is easily obtained by 
counting the reduction of the global symmetry 
$U(1)_B U(1)_L^3 U(1)_{PQ} U(1)_R
\rightarrow U(1)_B U(1)_L$ by introducing 
$\mu$ and SUSY breaking parameters.}. 
Most part comes from SUSY breaking 
parameters and especially from their flavor indices. 
Therefore, if these SUSY breaking parameters are arbitrary, 
MSSM induces too large flavor changing neutral current (FCNC) 
in $K^0- \overline{K^0}$ and $\mu \rightarrow e \gamma$, 
and too large 
electric dipole moment (EDM) of neutron and electron. 
Thus, if there exists SUSY in our world, 
there must exist underlying theory of SUSY breaking, 
which should naturally derive highly degenerated 
squark/slepton masses 
$\delta \tilde{m}^2 / \tilde{m}^2 < O(10^{-3})$ for the 
supersymmetric version of GIM cancellation to work, 
and 
small $CP$ phases (less than $O(10^{-2})$) 
in SUSY breaking parameters \footnote{A few TeV superparticle 
mass can also avoid large EDM.}.

\subsection{How is SUSY breaking messaged to our world?}

The underlying theory of SUSY breaking might be 
the model of spontaneous SUSY breaking. 
However, tree level supertrace formula 
${\rm Str} M^2 = 0$ 
shows that the squark/slepton which are lighter 
than quark/lepton 
appear when SUSY is spontaneously broken. 
Break through of this formula is possible by three ways. 
Correspondingly there are 
three ways to mediate SUSY breaking 
to our world, that is, \\
(a): Gravity mediate scenario (Hidden sector scenario) \cite{7}, \\
(b): Anomalous $U(1)$ gauge scenario \cite{8}, \\
(c): Gauge mediate scenario (Visible sector scenario) \cite{9}. \\
The 125(44) parameters in the MSSM 
decrease in each scenario. 
They are (a): 22(3), (b): 34(3) \footnote{Here we count anomalous $U(1)$ 
charges for the parameters.}, and 
(c): 22(3) \footnote{Here we count $\mu$ 2(1); 
$B$ 2(1). }. 
Each model has both merits and demerits. 
Here we concentrate on the scenario (c), which 
has the merits of calculability 
and naturally derives small FCNC since gauge interaction 
is flavor blind.

\subsection{A SUSY breaking model without messenger sector}

The dynamical SUSY breaking can explain 
the large hierarchy between the Planck scale 
and the weak scale. 
Eq.(\ref{univ}) shows that in the case of $N_f = N_c - 1$ 
the vacuum can be lifted dynamically. 
SUSY is broken in general when 
there are no no-flat directions 
and the continuous global symmetry is spontaneously 
broken \footnote{It is because 
the Goldstone boson can not have 
the supersymmetric partner\cite{NS}. }.
The 3-2 model \cite{3-2} is just the case
\footnote{The 3-2 model is the SM without 
the 
right-handed 
electron.}, whose 
superpotential is 
\begin{equation}
\label{W32}
 W_{3-2}  = \lambda X_1 + {2 \Lambda^7 \over X_3}. 
\end{equation}
The second term is induced from the non-perturbative effects 
as Eq.(\ref{univ}). 
$X_i$s $(i = 1 \sim 3)$ are moduli fields
\footnote{Three moduli fields remains in the Higgs phase: 
14 (degrees of freedom of fields) $-$ 
8 ($SU(3)$ gauge boson) $-$ 3 ($SU(2)$ gauge boson).}. 
Since we analyze the potential in the 
Higgs picture for the calculability, 
where gauge coupling is weak enough to use 
the canonical K\"ahler potential, 
$\lambda$ should be small of $O(10^{-7})$. 
Such small parameter can be avoided 
by building modified 3-2 model
\footnote{
It might be not so unnatural to 
introduce the scale $M_{\rm Planck}$ 
in the superpotential. 
It is because there exists supergravity theory 
behind. The massless Goldstone fermion can not 
be absorbed unless we consider 
supergravity theory.}$^{,}$
\footnote{$W_{3-2}'$ is obtained by changing the 
$R$ charge assignment of $X_{1,2}$. 
In this case $\lambda' \simeq O(1)$.} as 
\begin{equation}
\label{W32p}
 W_{3-2}'  = \lambda' {X_1 X_3 \over M_{\rm Planck}} + 
     {2 \Lambda^7 \over X_3}  .
\end{equation}

Now let us introduce the vector-like fields 
$q, \overline{q}$ and $l, \overline{l}$, 
which are the components of 
{\bf 5}, $\mbf{\bar{5}}$ of $SU(5)$. 
By the suitable $R$ charge assignment of 
vector-like fields, 
we can build a simple model of 
gauge mediation (c) without messenger sector \cite{Haba}, 
whose superpotential is 
\begin{eqnarray}
\label{W}
 W     &=& W_{3-2}' + h_1 
       {X_3 \over M_{\rm Planck}^3} q \overline{q} + 
       h_2 {X_3 \over M_{\rm Planck}^3} l \overline{l}\; . 
\end{eqnarray}
The flavor blind squark/slepton masses and gaugino mass 
are generated by loop diagram of gauge interactions. 
This is one of the simplest models of 
gauge mediation (c), 
which naturally avoid large FCNC since 
$\delta \tilde{m}^2 / \tilde{m}^2 \simeq 
\tilde{m}^2_{\rm gravity} / 
\tilde{m}^2_{\rm loop} \simeq O(10^{-4})$. 
The demerits of this model are that 
$\mu$ and $B$ are arbitrary parameters 
and that the color conserving minimum 
might not be the true vacuum \footnote{
We can easily show 
that there is no SUSY vacuum in this model.}.

\section{A composite model}

Next we would like to consider 
the application of the case $N_f = N_c + 1$. 
This is the case of confinement where 
only $M, B$, and $\overline{B}$ are massless fields 
bellow the strong dynamics scale. 
Here we consider enough high energy scale to 
neglect the SUSY breaking effects. 
People build composite models by using 
Eq.(\ref{1}) \cite{AN}$^{,}$ \cite{LM}. 
Here we show a model \cite{HO}$^{,}$ \footnote{
This model is the modified version of Ref \cite{LM}.}
which generates 
Yukawa coupling dynamically. 
Our motivation is very simple. 
It is that when we build the preon model 
whose ``meson'' $M$ is Higgs and 
``(ant-)baryon'' $B$ $(\overline{B})$ is 
quark/lepton, 
the first term of Eq.(\ref{1}) becomes 
the Yukawa coupling.

\par
We introduce $SU(7)_H$ gauge group and 
$N_f = 8$ preons $P_i, \overline{P_i}$, 
which also have 
SM gauge quantum numbers as shown in Table 1. 
When the $SU(7)_H$ gauge group becomes strong, 
confinement states $M$, $B$ and $\overline{B}$ 
appear as Table 2 and Table 3.

\begin{center}
\begin{table}[h]
\begin{center}
\begin{tabular}{|c||c|c|c|c||c|c|}
\hline
      & $SU(7)_H$ & $SU(3)_C$ & $SU(2)_L$ & $U(1)_Y$ &$U(1)_B$ & $U(1)_L$\\
\hline
\hline
$\pq$ & $\sb$  & $\sbb$ & $\sb$ & $-{1}/{3}$ & $-{1}/{21}$ & $ {2}/{7}$ \\
$\pl$ & $\sb$  &  $1$   & $\sb$ & $1$        & $ {2}/{7} $ & $-{5}/{7}$ \\
\hline
\hline
$\pu$ & $\sbb$ & $\sb$  & $1$   & $ {4}/{3}$  & ${1}/{21}$ & $-{2}/{7}$ \\
$\pd$ & $\sbb$ & $\sb$  & $1$   & $-{2}/{3}$  & ${1}/{21}$ & $-{2}/{7}$ \\
$\pn$ & $\sbb$ & $1$    & $1$   & $ 0 $       & $-{2}/{7}$ & ${5}/{7}$ \\
$\pe$ & $\sbb$ & $1$    & $1$   & $-2 $       & $-{2}/{7}$ & ${5}/{7}$ \\
\hline
\end{tabular}
\caption {Preons}
\end{center}
\end{table}
\end{center}

\vspace{-5mm}
 
\begin{table}[h]
\begin{center}
\begin{tabular}{|c||c|c|c|c||c|c||c|}
\hline
      & $SU(7)_H$ & $SU(3)_C$ & $SU(2)_L$ & $U(1)_Y$ &$U(1)_B$ &
 $U(1)_L$ & baryon $(\times 1/\Lambda^6)$\\
\hline
\hline
$Q$ & $1$ & $\sb$ & $\sb$ & ${1}/{3}$ & ${1}/{3}$ & $ 0 $ & ${\pq^5 \pl^2}$ \\
$l$ & $1$ & $1$   & $\sb$ & $-1$      & $ 0 $     & $ 1 $ & ${\pq^6 \pl}$ \\
\hline
\hline
$\bar{U}$ & $1$ & $\sbb$  & $1$ & $-{4}/{3}$ & $-{1}/{3}$ &
$0$&${\pu^2\pd^3\pe\pn}$ \\
$\bar{D}$ & $1$ & $\sbb$  & $1$ & $ {2}/{3}$ & $-{1}/{3}$ &
$0$&${\pu^3\pd^2\pe\pn} $ \\
$\bar{N}$ & $1$ & $1$     & $1$ & $ 0 $      & $ 0 $      &
$1$&${\pu^3\pd^3\pe} $ \\
$\bar{E}$ & $1$ & $1$     & $1$ & $ 2 $      & $ 0 $      &
$1$&${\pu^3\pd^3\pn} $ \\
\hline
\end{tabular}
\caption {``Baryon'' and ``anti-baryon''}
\end{center}
\end{table}
%
%
\begin{table}[h]
\begin{center}
\begin{tabular}{|c||c|c|c|c||c|c||c|}
\hline
      & $SU(7)_H$ & $SU(3)_C$ & $SU(2)_L$ & $U(1)_Y$ &$U(1)_B$ &
 $U(1)_L$ & meson $(\times 1/\Lambda)$\\
\hline
\hline
$\Phiu$  &$1$& $1$    &$\sb$&   $ 1$   &    $0$   & $0$&$\pq\pu$ \\
$\Phid$  &$1$& $1$    &$\sb$&   $-1$   &    $0$   & $0$&$\pq\pd$ \\
$\Gu$     &$1$&{\bf Ad}&$\sb$&   $ 1$   &    $0$   & $0$&$\pq\pu$ \\
$\Gd$     &$1$&{\bf Ad}&$\sb$&   $-1$   &    $0$   & $0$&$\pq\pd$ \\
$X$      &$1$& $\sbb$ &$\sb$&$-{1}/{3}$&$-{1}/{3}$& $1$&$\pq\pn$ \\
$\bar{X}$&$1$& $\sb$  &$\sb$&$ {1}/{3}$&$ {1}/{3}$&$-1$&$\pl\pd$ \\
$Y$      &$1$& $\sbb$ &$\sb$&$-{7}/{3}$&$-{1}/{3}$& $1$&$\pq\pe$ \\
$\bar{Y}$&$1$& $\sb$  &$\sb$&$ {7}/{3}$&$ {1}/{3}$&$-1$&$\pl\pu$ \\
$\phiu$  &$1$& $1$    &$\sb$&   $ 1$   &    $0$   & $0$&$\pl\pn$ \\
$\phid$  &$1$& $1$    &$\sb$&   $-1$   &    $0$   & $0$&$\pl\pe$ \\
\hline
\end{tabular}
\caption {``Meson''}
\end{center}
\end{table}
%

\noindent
Eq.(\ref{1}) is written by 
\begin{eqnarray}
& & W_{\rm dyn} = y_u\bar{U}(\Phiu + \phiu)Q  + 
                y_d\bar{D}(\Phid + \phid)Q  + 
                y_n\bar{N}(\Phiu + \phiu) \ell +
                y_e\bar{E}(\Phid + \phid) \ell  \nonumber  \\
            & &+ {\alpha_u} \bar{U} G_{\rm u} Q 
               + {\alpha_d} \bar{D} G_{\rm d} Q 
               + {\beta_u} \bar{U} \bar{Y} \ell
               + {\beta_d} \bar{D} \bar{X} \ell
               + {\beta_n} \bar{N}   X     Q
               + {\beta_e} \bar{E}   Y     Q    
               -  [ \mbox{Det} ] .
\end{eqnarray}
It shows that the Yukawa couplings are really 
generated dynamically. 
And this model predicts 4 Higgs doublets. 
Unfortunately, the color octet Higgs $G$ and 
lepto-quark $X, Y$ 
are also generated as ``meson'' fields, 
and 
$W_{\rm dyn}$ contains their Yukawa couplings. 
Since they are all massless, 
we must introduce 
\begin{eqnarray}
\label{4}
        W_{tree} = g_{G}   {{\pq\pq\pu\pd}\over{M}}
                 + g_{\mu} {{\pl\pl\pe\pn}\over{M}}
                 + g_{x}   {{\pq\pl\pd\pn}\over{M}}
                 + g_{y}   {{\pq\pl\pu\pe}\over{M}} 
\end{eqnarray}
by hand in order to give them masses. 
Eq.(\ref{4}) is the most general 
gauge invariant 4 point interactions, 
which also generate 
$\mu$ terms ($\mu \simeq \Lambda^2/M$). 
\par
This model \footnote{The anomalies in this model are the same 
in the SM, 
$SU(2)_{L}^2 U(1)_{B+L}, U(1)_{Y}^2 U(1)_{B+L}$.} 
can be extended to the 3 generation model. 
Considerable extensions are 
(1): $SU(7)_H \rightarrow SU(7)_H^3$, 
(2): $SU(7)_H \rightarrow SU(23)_H$ \footnote{It 
corresponds to the extension of global symmetry 
$U(1)_B U(1)_L \rightarrow SU(3)_F 
U(1)_B U(1)_L$.}, 
(3): The first and second generations are elementary 
and the third generation and Higgs are 
composite \footnote{ 
The idea (3) comes from the fact that preons 
(Table 1) are just like ``mirror fields''. 
Mirror fields are conjugated states of ordinal fields, 
which appear in string theory.}.

\section{Summary}

\par
We have shown two models of phenomenological 
application of the non-perturbative effects in SUSY QCD. 
One is the low energy SUSY breaking scenario 
without messenger sector, 
and the second is the composite model generating 
Yukawa interaction dynamically. 
It is interesting to try to apply the non-perturbative 
effects in SUSY QCD to other phenomenologies. 
\par
\section*{Acknowledgments}
It is my pleasure to thank the organizers of 
PPPP'97 (APCTP) for their 
kind hospitality and for giving me 
the opportunity to present this work. 
I would like to thank T. Matsuoka, N. Okamura and N. Maru 
for the collaboration on these works. 
I would also like to thank 
K. Hagiwara for continuous encouragements.

\end{document}